\documentclass[aps,pra,twocolumn,showpacs,superscriptaddress]{revtex4-1}
\usepackage{enumerate}
\usepackage{graphicx}
\usepackage{amsmath}
\usepackage{amsfonts}
\usepackage{amssymb}
\usepackage{braket}
\usepackage{gnuplottex}
\usepackage{epsf}

\begin{document}
\today
\title{Validity of rotating wave approximation in non-adiabatic holonomic quantum 
computation}
\author{Jakob Spiegelberg}
\affiliation{Department of Quantum Chemistry, Uppsala University,
Box 518, Se-751 20 Uppsala, Sweden}
\author{Erik Sj\"oqvist}
\affiliation{Department of Quantum Chemistry, Uppsala University,
Box 518, Se-751 20 Uppsala, Sweden}
\affiliation{Centre for Quantum Technologies, National University
of Singapore, 3 Science Drive 2, 117543 Singapore, Singapore}
\date{\today}
\begin{abstract}
We examine the validity of the rotating wave approximation (RWA) in non-adiabatic holonomic 
single-qubit gates [New J. Phys. {\bf 14}, 103035 (2012)]. We demonstrate that the adoption of 
RWA may lead to a sharp decline in fidelity for rapid gate implementation and small energy 
separation between the excited and computational states. The validity of the RWA in the recent 
experimental realization [Nature (London) {\bf 496}, 482 (2013)] of non-adiabatic holonomic 
quantum computation for a superconducting qubit is examined. 
\end{abstract}
\maketitle
Holonomic quantum computation (HQC) is the idea of using non-Abelian geometric phases to 
implement robust quantum gates \cite{zanardi99}. By using adiabatic holonomies, HQC becomes 
tolerant to errors caused by fluctuations of the slowly changing control parameters. On the 
other hand, dissipation may have detrimental effects on the gates, leading to the need of 
performing the gate operations as fast as possible by using non-adiabatic holonomies.  
Non-adiabatic strategies have been shown \cite{johansson12} to be effective to minimize 
this error source. However, a shortening of the run time may in turn lead to other errors that can 
lower the gate fidelity and therefore put a limitation on the speed of holonomic gate operations. 
Here, we examine how the validity of the rotating wave approximation (RWA) depends on the 
run time and energy structure of the three level $\Lambda$ setting used to implement 
non-adiabatic non-Abelian geometric gates first proposed in Ref. \cite{sjoqvist12} and 
experimentally demonstrated in Refs. \cite{abdumalikov13,feng13}. 

The speed of quantum gate operations is generally limited by unwanted effects that become 
more pronounced when the run time is decreased. One such effect is related to the  
quasi-monochromatic approximation \cite{mandel95} that breaks down for short pulses, 
causing population of energy levels outside the computational subspace \cite{steane00,berger12}. 
Another speed-limiting feature is the RWA, which is expected to break down when the run time of 
the gate becomes too short. This leads to a situation where the Rabi flopping is accompanied 
by faster fidelity reducing oscillations \cite{shahriar04}; an effect that can be suppressed by 
embedding the qubit in an off-resonant $\Lambda$ system \cite{pradhan09}. We quantify 
the validity of the RWA by computing the fidelity of the ideal RWA-based non-adiabatic holonomic 
single-qubit gate operations with respect to numerical solutions of the exact Schr\"odinger equation. 

The $\Lambda$ system consists of states $\ket{0}$ and $\ket{1}$ coupled to the 
auxiliary excited state $\ket{e}$ via a pair of oscillating electric field pulses ${\bf E}_j (t) = 
\boldsymbol{\epsilon}_j g_j(t) \cos (\omega_j t)$, $j=0,1$, $g_j (t)$ being envelope functions 
describing the pulse shape and duration. The polarization $\boldsymbol{\epsilon}_j (t)$ is 
chosen so as to allow for the $j \leftrightarrow e$ transition only. The Hamiltonian of the 
system reads $\hat{H} (t) = \hat{H}_0 + \hat{\boldsymbol{\mu}} \cdot \left[ {\bf E}_0 (t) + 
{\bf E}_1 (t) \right]$, where $\hat{H}_0 = - f_{e0} \ket{0} \bra{0} - f_{e1} \ket{1} \bra{1}$ is 
the bare Hamiltonian (by putting the energy of the excited state to zero) and 
$\hat{\boldsymbol{\mu}}$ is the electric dipole operator. By tuning the oscillation frequencies 
$\omega_j$ on resonance with the bare transition frequencies $f_{ej}$, the Hamiltonian 
in the interaction picture reads 
\begin{eqnarray}
\hat{H}_I (t) & = & \Omega_0(t) \left(1+e^{-2if_{e0}t}\right)\ket{e}\bra{0} 
\nonumber \\ 
 & & + \Omega_1(t) \left(1+e^{-2if_{e1}t}\right)\ket{e}\bra{1} + 
{\textrm{h.c.}} , 
\end{eqnarray}
where $\Omega_j (t) = \bra{e} \hat{\boldsymbol{\mu}} \cdot \boldsymbol{\epsilon}_j \ket{j} 
g_j (t) /(2\hbar)$. The RWA means that the $e^{\pm2if_{ej} t}$ terms oscillate rapidly 
enough so that they can be neglected in $\hat{H}_I (t)$.  

Provided the RWA applies, a non-adiabatic holonomic gate $\hat{U}(C)$ acting on the 
computational subspace spanned by $\ket{0}$ and $\ket{1}$ is implemented by choosing 
electric field  pulses such that $\Omega_0 (t) / \Omega_1 (t)$ is time independent and the 
$\pi$ pulse criterion $\int_0^{\tau} \sqrt{|\Omega_0 (t)|^2 + |\Omega_1 (t)|^2} dt = \pi$ is 
satisfied. Here, the latter condition assures that the computational subspace undergoes a 
cyclic evolution around the path $C$ in the Grassmannian $G(3;2)$ \cite{bengtsson06}; the 
former guarantees that the dynamical phases vanish for the full pulse duration, which 
implies that the gate is fully determined by $C$ \cite{sjoqvist12}. Explicitly, 
\begin{eqnarray}
\hat{U}(C) = \sin \theta \cos \phi \hat{\sigma}_x + \sin \theta \sin \phi \hat{\sigma}_y + 
\cos \theta \hat{\sigma}_z ,
\end{eqnarray} 
where $\Omega_0 (t) / \Omega_1 (t) = -\tan (\theta /2) e^{i\phi}$ and $\hat{\sigma}_k$, 
$k=x,y,z$, are the standard Pauli operators acting on the computational subspace. Here, 
we examine the gate fidelity by comparing the ideal holonomic RWA-based transformation 
$\ket{\psi} \mapsto \hat{U}(C) \ket{\psi}$ with the transformation $\ket{\psi} \mapsto 
{\bf T} e^{-(i/\hbar) \int_0^{\tau} \hat{H}_I (t) dt} \ket{\psi} $ obtained by numerically 
solving the exact Schr\"odinger equation for $\hat{H}_I (t)$, in order to determine the range 
of validity of the RWA. The gate fidelity for an input state $\ket{\psi}$ is given by 
$\left| \bra{\psi} U^{\dagger} (C) {\bf T} e^{-(i/\hbar) \int_0^{\tau} \hat{H}_I (t) dt} 
\ket{\psi} \right|$, i.e., the overlap between the exact and the ideal RWA-based 
outputs. ${\bf T}$ denotes time ordering. 

In order to test the validity of the RWA, the dependence of the fidelity on transition frequencies, 
pulse shape, and pulse duration is examined. As holonomic test gates, we choose the NOT gate 
$\ket{x} \mapsto \ket{x\oplus 1}$, where $x=0,1$ and $\oplus$ is addition modulo $2$, which 
is achieved in the RWA regime by setting $\Omega_{0} (t) / \Omega_{1} (t) = 1$, and the Hadamard 
gate $\ket{x} \mapsto \frac{1}{\sqrt{2}} ((-1)^x \ket{x} + \ket{x\oplus 1})$, where 
$\Omega_{0} (t) / \Omega_{1} (t) = -\tan \left( \frac{\pi}{8} \right)$. 

\begin{table}
\centering
\begin{tabular}{ccc} \hline
$f$ [$s^{-1}$] \,  \,  \, & NOT & \, Hadamard  \\ \hline
$10^6$          & 0.0037 & 0.7071   \\
$10^7$          & 0.0394 & 0.7004   \\
$10^8$          & 0.8543 & 0.7903   \\
$5 \times 10^8$  & 0.9750 & 0.9712  \\
$10^9$          & 0.9990 & 0.9994  \\
$10^{10}$       & 1.0000 & 1.0000  \\ \hline   
\end{tabular}
\caption{Fidelity of holonomic NOT and Hadamard gates for different transition frequencies
$f \equiv f_{e0} = f_{e1}$. We use a truncated Gaussian shaped pulse with full width at half 
maximum (FWHM) = 10 ns, a total duration of 40 ns, and input state $\ket{0}$.}
\label{transition}
\end{table}

In Table \ref{transition}, fidelities for a range of transition frequencies $f \equiv f_{e0} = f_{e1}$ 
are displayed for the two gate operations acting on the input state $\ket{0}$. In both cases, three 
different regions can be identified: For large $f_{ej}$, the RWA is valid and the exact and ideal 
output states nearly coincide leading to a fidelity close to unity. For small energy separations, 
the additional exponential term leads to a factor of 2 since $1+e^{2if_{ej}t} \approx 2$. The 
quantum system runs the cyclic evolution twice. One property of the matrices representing 
Hadamard and NOT gate is that their product with themselves is the identity matrix, the new 
transformation resembles the identity operation that preserves the input state. In case of the 
NOT gate, the overlap between the input state $\ket{0}$ and the output state after running 
through a NOT gate vanishes per definition. For the Hadamard gate, the overlap between input 
state $\ket{0}$ and Hadamard transformed state is $1/\sqrt{2} \approx 0.7071$. In the third 
region, between these extremes, the RWA leads to oscillations of the overlap. The impact on 
the system would be highly dependent upon the precise timing of the laser pulses and the 
corresponding operation would therefore not represent a simple quantum gate in this region. 
These findings still hold for a situation where the transition frequencies are not equal, i.e., 
$f_{e0} \neq f_{e1}$. 

\begin{table}
\centering
\begin{tabular}{cccc} 
\hline
\, Envelope & \, $\frac{1}{\sqrt{2}} (\ket{0} + \ket{1})$ \, & \,  
$\frac{1}{\sqrt{2}} (\ket{0} + i\ket{1})$ \, & $\ket{0} $ \\ 
\hline
Truncated Gaussian & 0.9999 & 0.9853 & 0.9861 \\
sech & 0.9956 & 0.9953 & 0.9947 \\
Parabola & 0.9991 & 0.9988 & 0.9988  \\
sin$^2$ & 0.9975 & 0.9962 & 0.9959  \\ 
Square & 0.9991 & 0.9989 & 0.9980 \\ 
\hline
\end{tabular}
\caption{Fidelity of holonomic NOT gate for different envelope functions and the $+1$ 
eigenstates of $\hat{\sigma}_x,\hat{\sigma}_y,\hat{\sigma}_z$, respectively, as input states. 
All other system parameters are taken from Ref. \cite{abdumalikov13}.}
\label{shape1}
\end{table}

\begin{table}
\begin{tabular}{ccccc} \hline
\, Envelope \, &  100ns  &  40ns  &  10ns  &  2.5ns  \\ \hline 
Truncated Gaussian & 0.9987 & 0.9861  & 0.8072  & 0.1790   \\   
sech & 0.9995 & 0.9947  & 0.9792  & 0.6703   \\
Parabola & 0.9997 & 0.9988  & 0.9987  & 0.8573   \\
sin$^2$ & 0.9996 & 0.9959  & 0.9857  & 0.4424         \\
Square & 0.9998 & 0.9980  & 0.9991  & 0.7952   \\ \hline
\end{tabular}
\caption{Fidelity of the input state $\ket{0}$ after a NOT transformation for a selection of total 
durations of the pulses. All other system parameters are taken from Ref. \cite{abdumalikov13}.}
\label{shape2}
\end{table}

Next, five different pulse shapes are tested for the $+1$ eigenstates $\ket{0}, 
\frac{1}{\sqrt{2}} (\ket{0} + \ket{1}),\frac{1}{\sqrt{2}} (\ket{0} + i\ket{1})$ of the three Pauli 
operators $\hat{\sigma}_x,\hat{\sigma}_y,\hat{\sigma}_z$ as input. A truncated 
Gaussian pulse with the full width at half maximum (FWHM) as one fourth of the full pulse 
duration, a secant pulse, a parabolic pulse, a sin$^2$ pulse, and a square pulse. The fidelities 
for the respective cases are enlisted in Table \ref{shape1}. There are only small differences in 
the fidelity for different pulse shapes. However, the truncated Gaussian pulse leads to comparably 
low fidelities. An explanation for this deviation can be found in Table \ref{shape2}. Since the 
FWHM of the truncated Gaussian pulse is chosen as one fourth of the absolute pulse duration, 
the region where the envelope is significantly different from zero is in the same order of 
magnitude as the FWHM. We suspect that only these regions contribute significantly to the 
system dynamics. Hence, the truncated Gaussian pulse for 40 ns is comparable to a 10 ns 
pulse of the other shapes. 

\begin{figure}[h]
\includegraphics[scale=0.3]{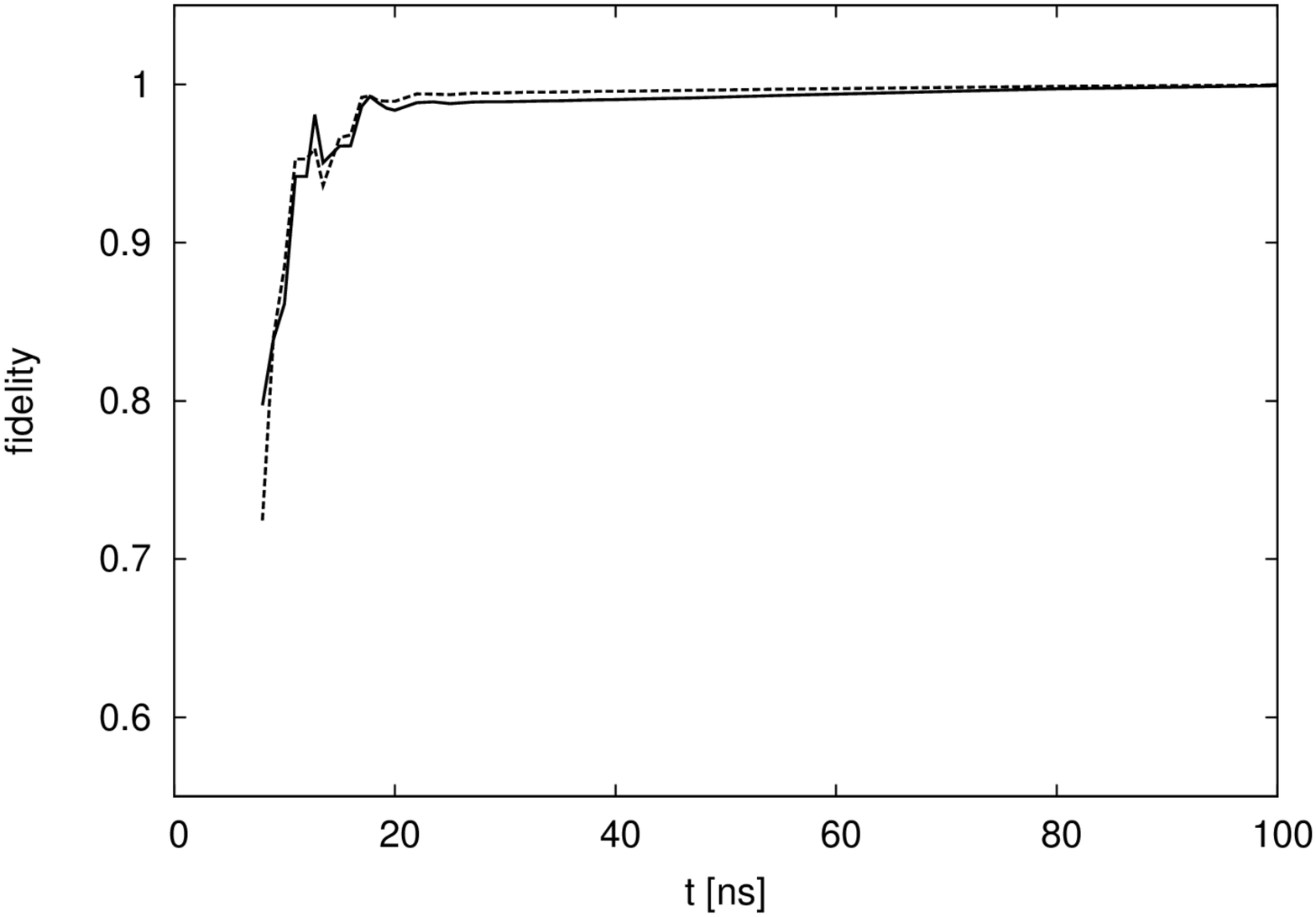}
\caption{Dependence of the fidelity on the pulse duration for the NOT gate (continuous line) and the 
Hadamard gate (dashed line). A truncated Gaussian pulse is used. All other system parameters 
are taken from Ref. \cite{abdumalikov13}.}
\label{duration}
\end{figure}

Very short laser pulses have the advantage that dissipation can be neglected. However, the RWA 
leads to instabilities in the quantum gate, if the time scale becomes too short. The fidelity achieved 
with a truncated Gaussian laser pulse as a function of the total pulse duration is shown in Fig. 
\ref{duration}. The calculation is based on averaging the output state overlap over the input 
states $\ket{0},\frac{1}{\sqrt{2}} (\ket{0} + \ket{1}),\frac{1}{\sqrt{2}} (\ket{0} + i\ket{1})$. For 
both gates examined, the fidelity is stable over a wide range. Upon some threshold, the fidelity 
begins to fluctuate heavily and deteriorates quickly, in this case at approximately 20 ns. 

\begin{figure}[h]
\includegraphics[scale=0.3]{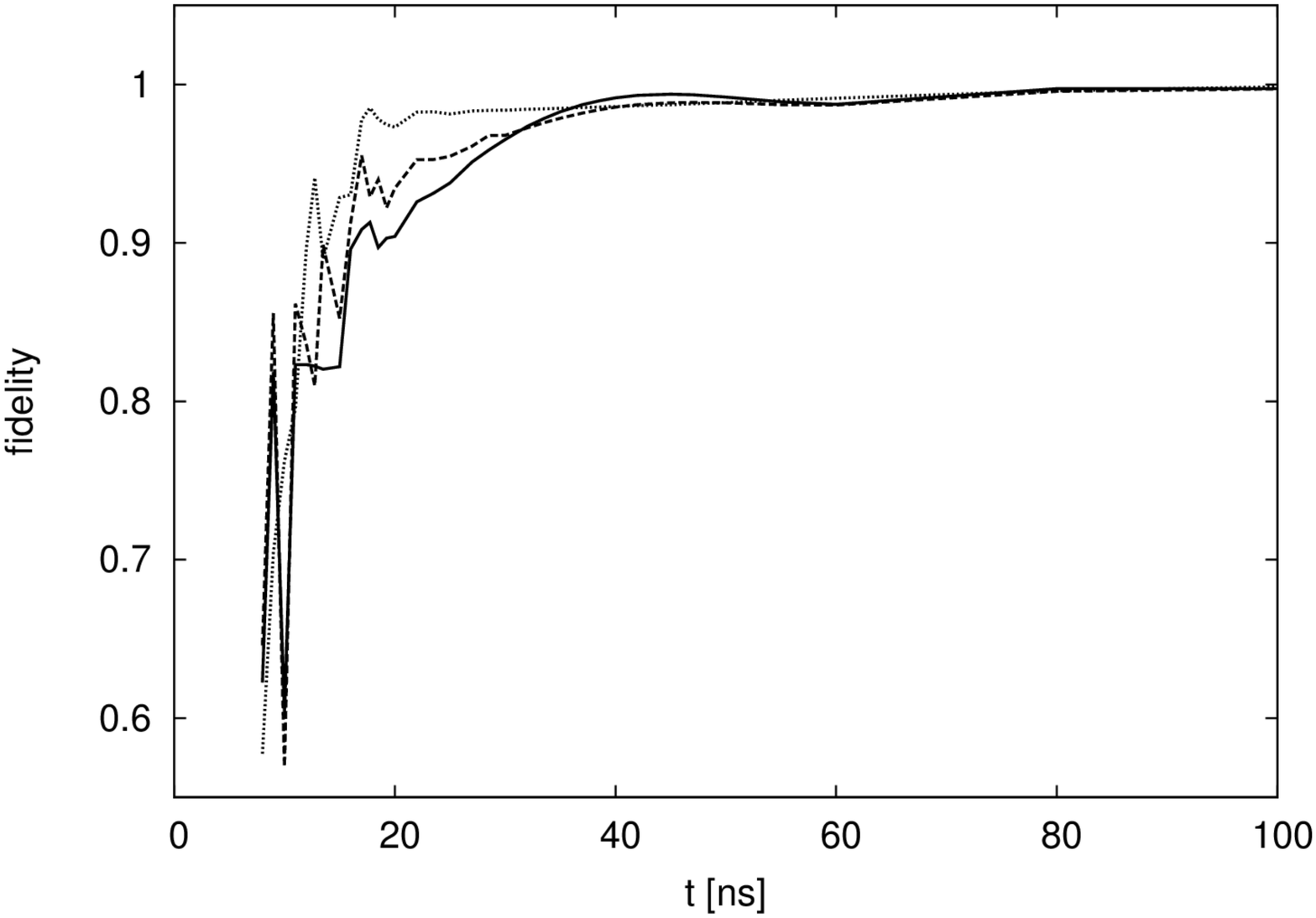}
\caption{Dependence of the fidelity on the pulse duration for the combinations of Hadamard and 
NOT gate (continuous line), NOT and Hadamard gate (dashed line) and the product of the fidelities 
of the NOT and Hadamard gates separately (dotted line). The duration plotted here equals the 
duration of each of the pulses and thereby half of the combined gate's duration. A truncated 
Gaussian pulse is used. All other system parameters are taken from Ref. \cite{abdumalikov13}.}
\label{consecutive}
\end{figure}

Abdumalikov {\it et al.} \cite{abdumalikov13} experimentally realized gate operations on a 
$\Lambda$ system implemented in a transmon superconducting qubit, following the proposal 
in Ref. \cite{sjoqvist12}. Their transition frequencies were $f_{e0} \approx 5.0806 \times 10^{10}$ 
s$^{-1}$ and $f_{e1} \approx 4.8580 \times 10^{10}$ s$^{-1}$; their truncated Gaussian shaped 
pulse had a FWHM of 10 ns and a full duration of 40 ns. This choice of parameters lies relatively 
close to the edge of the zone with stable fidelity. If a further speed up of the computation should 
be achieved, a larger separation of the energy levels has to be aimed for.

In holonomic quantum computing, non-commuting gates can be implemented. Here, we studied 
the influence of the RWA on the fidelity after application of the two possible combinations of NOT 
and Hadamard gate and compared these with the fidelity obtained by multiplying the fidelities for 
separate NOT and Hadamard gate, i.e., the fidelity as if the two gates were commuting. In our model, 
the two pulses followed each other without any separation in time.

As can be seen in Fig.~\ref{consecutive}, the fidelity achieved with the non-commuting gates
is typically lower than the product of their fidelities. Furthermore, the fidelity decreases at pulse 
durations around 40 ns. This decline appears earlier than expected taking the multiplied gate 
fidelities of the separate gates as reference.  We conclude that this effect occurs due to the 
non-Abelian nature of the gates. The fluctuations which were previously only significant in the 
short duration regime, are now visible throughout the entire range of durations studied. The 
non-commutivity of Hadamard and NOT gate can be clearly seen by comparison of the H-NOT 
and NOT-H combinations in Fig.~\ref{consecutive}.

To conclude, the rotating wave approximation (RWA) has been proven to be valid in the three-level 
setup designed for non-adiabatic holonomic quantum computation proposed in Ref. \cite{sjoqvist12} 
over a wide range system parameters. Only at small transition frequencies and very fast pulses, the 
RWA has an impact on the quantum systems evolution. The order of magnitude of state energy 
separation in atomic or molecular systems lies typically above this problematic region, already 
a separation of several meV is sufficient. Possible problems arising through the RWA can thus 
be avoided.
\vskip 0.3 cm
E.S. acknowledges support from the National Research Foundation and the Ministry of 
Education (Singapore).

\end{document}